\documentclass[epsf,twocolumn,showpacs,preprintnumbers]{revtex4}
\usepackage{graphics}
\usepackage{graphicx}
\usepackage{dcolumn} 
\usepackage{bm}
\usepackage{epsfig}
\begin{document}
\title{Correlation Effects in the 
  Triangular Lattice Single-band System Li$_x$NbO$_2$} 
\author{K.-W. Lee$^1$, J. Kune\v{s}$^{2,3}$, R. T. Scalettar$^1$, 
     and W. E. Pickett$^1$} 
\affiliation{$^1$Department of Physics, University of California, Davis, 
 California 95616, USA}
\affiliation{$^2$Theoretical Physics III, Center for Electronic Correlations
 and Magnetism, Institute of Physics, University of Augsburg,
 Augsburg 86135, Germany}
\affiliation{$^3$Institute of Physics, Academy of Sciences,
   Cukrovarnick\'a 10, CZ-162 53 Prague 6, Czech Republic}
\date{\today}
\begin{abstract}
Superconductivity in hole-doped Li$_x$NbO$_2$ has been reported with 
$T_c \approx$5 K in the range 0.45 $\le x<$ 0.8.
The electronic structure is based on a two-dimensional
triangular Nb lattice.
The strong trigonal crystal field results in a single Nb $d_{z^2}$
band isolated within a wide gap, leading to 
a single-band triangular lattice system.
The isolated, partially filled band has a width $W$=1.5 eV with dominant
{\it second} neighbor hopping.
To identify possible correlation effects, we apply DMFT(QMC) using on-site
Coulomb repulsion $U$=0--3 eV, and check selected results using
determinant QMC.
For $U$ as small as 1 eV, the single particle spectrum 
displays a robust lower Hubbard band, suggesting 
the importance of correlation effects in Li$_x$NbO$_2$ 
even for $U \le W$.
At half-filling ($x$=0), a Mott transition occurs
at $U_c \approx$1.5 eV.
Coupling between the O $A_g$ phonon displacement and correlation
effects is assessed. 
\end{abstract}
\maketitle

\section{Introduction}
Not long after the discovery of high temperature superconductivity
in layered cuprate compounds, Geselbracht, Richardson, and Stacy 
discovered superconductivity at $x$=0.45 and 0.50 in the layered
niobate Li$_x$NbO$_2$ with $T_c \approx$5 K.\cite{gese1}
Samples were obtained by removing Li (hence electrons) from band
insulating LiNbO$_2$.
This superconductivity has been confirmed in the range 0.45$\le x\le$0.79
with $T_c$ showing no significant dependence on
Li concentration,\cite{gese1,bordet,mosh,liu} 
but no superconductivity in the weakly doped range between 
$x$=0.84 and $x$=1.\cite{mosh}
Hall effect measurements confirmed hole-type carriers,\cite{keller1}
consistent with LiNbO$_2$ being a band insulator.
The difficulty in synthesizing materials with $x<$0.45
may be related to the considerable
covalency of Li along $c$-direction that is reflected in the Born 
effective charge.\cite{erik1,erik2}

Recently, Liu {\it et al.} revisited the superconductor ($x$=0.68)
using specific heat measurements.\cite{liu}
The electronic specific heat can be
fit well using the BCS-type $s$-wave symmetry pairing.
Additionally, the linear specific heat coefficient 
$\gamma_{exp}$=3.59 mJ/mol K$^2$ and the Debye temperature 462 K
were obtained.
The virtual crystal approximation (Li nuclear charge $Z=2+x$),
leads to the band structure value $\gamma_{b}$=2.425 mJ/mol K$^2$ at $x=0.68$,
corresponding to weak electron-phonon coupling strength
$\lambda = \frac{\gamma_{exp}}{\gamma_{b}}-1 \approx$0.48, which could
however be enough to account
for $T_c \approx 5$ K.
The electron-phonon mechanism is consistent with stated theoretical  
viewpoints.\cite{erik1,freeman} 

There are two interesting sister materials.
An isostructural and isovalent Na$_x$NbO$_2$ has been observed
to be superconducting with a little lower $T_c \approx$4 K,
in the range $0.5 \le x \le 0.7$.\cite{keller2,keller3}
Interpreted as an isotope shift, $\alpha=-d(ln T_c)/d(ln M)$ (M=mass) is
0.27. This value might be thought to imply phonon-mediated pairing, but 
the important coupling is not expected to be to the Li (Na) ion.
Another system is H$_x$LiNbO$_2$, showing
superconductivity with $T_c$=5 K at $x$=0.3 and 0.5.\cite{kumada2}
Due to lack of structure studies as well as doping studies,
the mechanism and what role H plays are not yet resolved.

In the ${\cal M}$(S,Se)$_2$ system with a transition metal ${\cal M}$,
competition between a charge density wave transition (CDW) 
and superconductivity is frequently observed.
Recently, Morosan {\it et al.} synthesized Cu$_x$TiSe$_2$.\cite{tise2} 
By intercalating Cu (i.e., electron donation from Cu), 
the CDW is suppressed and this system becomes 
superconducting with the maximum $T_c \approx$4 K at $x=0.08$.
The Wilson ratio $R \sim 0.35$ is small, implying
the importance of correlation effects.
Another compound for comparison is Na$_x$CoO$_2$, which becomes
a superconductor at $x \sim 1/3$ by intercalating water.
Also, the dehydrated compound displays an interesting phase diagram 
with unique insulating phase at $x=0.5$ and
shows large correlation effects. 
As intensively studied by some of the present authors,\cite{nacoo}
correlation effects in this system lead to charge disproportionation on 
the Co sublattice.

In addition to the common layered structure, both ${\cal M}$(S,Se)$_2$ and
Cu$_x$TiSe$_2$ have 
an edge-sharing octahedral framework,
similar with NbO$_6$ trigonal prismatic coordination in Li$_x$NbO$_2$.
These similarities imply the significance of correlation effects
in this niobate. 
At $x=0$ (NbO$_2$) Nb has only one valence electron ($4d^1$)
and the 
possibility of a Mott transition naturally arises.  The observed 
structure of NbO$_2$ is distorted rutile,\cite{nbo2} and the layered
form may not be accessible experimentally.

In this paper, we investigate correlation effects
on the presumed Mott insulating and nonstoichiometric phases
using the dynamical mean field theory approximation (DMFT).\cite{dmft}
Selected results are compared with determinant quantum Monte Carlo 
calculations, where intersite correlations are included and the only
limitations are the lattice size and the temperature.
For varying strength of repulsion (whose value is not known), and for
fillings $n$=1, 4/3, and 5/3 (note $n=x+1$) we obtain the spectral
density, and at $n$=1 we identify the critical strength for a metal-
insulator transition, and display the evolution of system characteristics
through the transition.   Finally, we present changes in the spectral
density due to (frozen) $A_g$ phonon displacements. 

The triangular lattice aspect itself has assumed renewed interest,
due in part to the discovery of superconductivity in the Na$_x$CoO$_2$
system.  The nearest-neighbor triangular Hubbard model has been studied to
uncover how it differs from the square lattice that is used as a model of
the high temperature superconductors; its non-bipartite nature is found
to be important in frustrating local spin correlations.  We will compare
results for Li$_x$NbO$_2$ where 2nd neighbor coupling dominates, to 
the same dynamical mean field approximation calculations for the
nearest-neighbor case of Aryanpour and collaborators.\cite{karan}  

\section{Structure and calculation}
\label{method}
\begin{figure}[tbp]
\rotatebox{-90}{\resizebox{6.5cm}{7.5cm}{\includegraphics{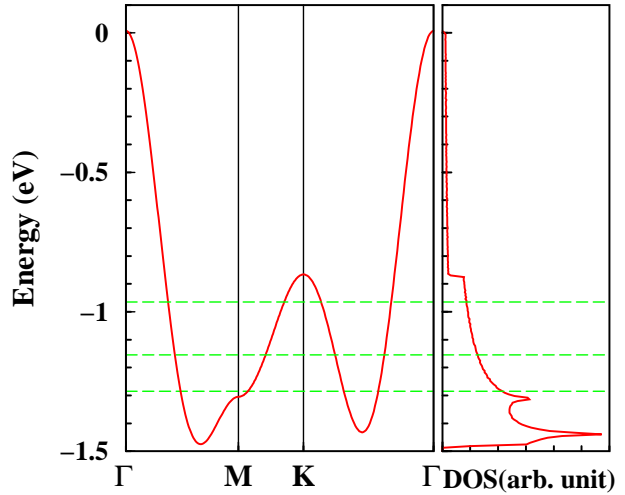}}}
\caption{(Color online) Two-dimensional tight-binding band structure and
 density of states, using up to third neighbor hopping parameters
 ($t_1$=64, $t_2$=100, and $t_3$=33 meV), for the single isolated 
 $d_{z^2}$ state in LiNbO$_2$.
 The dashed horizontal lines indicate the Fermi energy $E_F$ 
 in the rigid band picture for electron fillings $n$=1, 4/3, 
 and 5/3 from bottom to top.
 The centroid of the band $E_0$ lies on 18 meV below $E_F$ 
 for $n$=4/3.
Note lower half of weight is localized within 0.2 eV width, only
 13 \% of the total bandwidth.
}
\label{band}
\end{figure}

 LiNbO$_2$ is based on a double-layered hexagonal structure 
($P6_3/mmc$, No.~194) with lattice parameters $a$=2.90~\AA~and 
$c$=10.46~\AA.\cite{bordet, meyer,kumada,gese2,mosh,cherk}
Li and Nb atoms lie on $2a$ sites (0,0,0) and $2d$ sites 
($\frac{2}{3}$,$\frac{1}{3}$,$\frac{1}{4}$), respectively.
O atoms occupy $4f$ sites ($\frac{1}{3}$,$\frac{2}{3}$,$z$).
In this material, NbO$_6$ units place the Nb ion in
trigonal prismatic coordination. 
The O internal parameter $z$, which is measured from the Li layers,
shows some disagreement between reports, and our calculations 
use the experimental value $z$=0.1263\cite{meyer} for which the electronic
structure was analyzed earlier.\cite{erik1} 
This value leads to Nb-O bond length 2.116~\AA, and O-Nb-O bond angles
86.5$^\circ$ and 75.4$^\circ$. 

We have carried out first principles, local density approximation
(LDA)  calculations using the full-potential 
local-orbital code (FPLO),\cite{fplo} and the results 
are consistent with previous reports.\cite{erik1,erik2}
The basis set was Li $(1s)2s2p3d$, Nb $(4s4p)5s5p4d$,
and O $2s2p3d$. (The orbitals in parentheses indicate semicore 
orbitals.)
As reported previously,\cite{erik1,freeman} the electronic structure shows
strong two-dimensionality 
and for our investigation of correlation effects we neglect 
the $k_z$ dispersion.
Our DMFT calculations were carried with two-dimensional
tight-binding (TB) parameters $t_1$=64, $t_2$=100, and $t_3$=33 meV,
leading to bandwidth $W$=1.5 eV.
These hopping parameters were obtained using Wannier function 
techniques.\cite{erik1}
The TB band structure and DOS are displayed in Fig. \ref{band}.

Studies of lithium-deintercalation in the range 
$0.5\le x \le 1$ by Kumada {\it et al.} display negligible
change in the structure parameters including Nb-O bond length.\cite{kumada}
This behavior is consistent with the observation that 
the electronic structure follows a rigid band model 
well.\cite{erik1,freeman,cherk}
Thus these tight-binding parameters will be used for all $x$.
(Since another structure\cite{nbo2} is observed, $2H$--NbO$_2$ as 
considered here would be at best a metastable state.)

Hirsch-Fye quantum Monte Carlo (QMC) DMFT calculations\cite{dmft} were  
carried out
to investigate dynamic correlation
effects, applying the on-site Coulomb repulsion $U$ less than 
3 eV to the single isolated $d_{z^2}$ Wannier function.
The DMFT study is based on the Hubbard model
\begin{eqnarray}
H &=& \sum_{i,j, \sigma}t_{ij} 
   (c_{i\sigma}^+c_{j\sigma} + c_{j\sigma}^+c_{i\sigma}) 
  -\mu \sum_{i\sigma}n_{i\sigma} \nonumber\\
  &+& U\sum_{i}n_{i\uparrow}n_{i\downarrow}
\end{eqnarray}
with the hopping parameters $t_{ij}$ between site $i$ and $j$, 
the chemical potential $\mu$, the spin quantum number 
$\sigma$, $c_{j\sigma}^+$ creates an electron of spin $\sigma$ at site $j$,
and $n_{i\sigma}\equiv c_{i\sigma}^+c_{i\sigma}$ is the number operator.
The results reported here are for $T$=1100 K (i.e., 0.1 eV).
40--100 time-slices are used for the QMC runs.

The spectral density $A(\omega)$ has been obtained using the maximum-entropy
(MaxEnt) analytic continuation technique developed by Jarrell and
Gubernatis.\cite{maxent}
In MaxEnt good QMC data, close to Gaussian distribution, 
are crucial.
We checked the statistics up to 12$\times 10^6$ sweeps and also
about 500 $k$-points in the two-dimensional irreducible wedge.
However, at small and large $U$ regimes, the spectral densities,
specially near $\omega=0$, depend noticeably on the convergence parameters.
In this paper we will show the results in range of $U$=0.5--2.5 eV, 
which shows robust results for $A(\omega)$.
MaxEnt can be performed in a few different ways. In our calculations,
both classic and Bryan techniques have been used and show consistent 
results.
The other ingredient in MaxEnt is the default model for the
spectral density.\cite{maxent}
Since the default model is immaterial for good data, 
two different ways based on a Gaussian model were tried.
One is independently calculated using the same default model 
for each $U$, and the other way is that a converged result for
previous higher $U$ value is used as a default model 
for next small $U$.
These methods give consistent results, indicating the robustness of results
from the MaxEnt technique.

The bulk of our results have been obtained with this DMFT approach.
To assess whether intersite correlations may affect our conclusions,
we complement them with a limited set of simulations using determinant
quantum Monte Carlo (DQMC) \cite{bss}, performed for the same
Hamiltonian, Eq.~1 but on finite spatial lattices.  This approach
includes the momentum dependence of the self energy, but has the
drawback of not working in the thermodynamic limit, as DMFT does.
As in the DMFT QMC solver, DQMC requires a discretization of the
inverse temperature $\beta$.  We have chosen the interval
$\Delta \tau = 0.5$ which satisfies the condition
$t_i U (\Delta \tau)^2 << 1$ required for the `Trotter errors'
associated with the discretization to be small.  Our runs
typically used 2,000 sweeps to equilibrate the lattice and
20,000 sweeps to take measurements.  We have looked at the density
and short range spin correlations on 12$\times$12 lattices in addition to
the 9$\times$9 lattice results shown here, and observe only minor changes.

\section{Metal-Insulator Transition}
\label{MIT}
\subsection{Electron Filling versus Chemical Potential}
\begin{figure}[tbp]
\rotatebox{-90}{\resizebox{6.5cm}{7.5cm}{\includegraphics{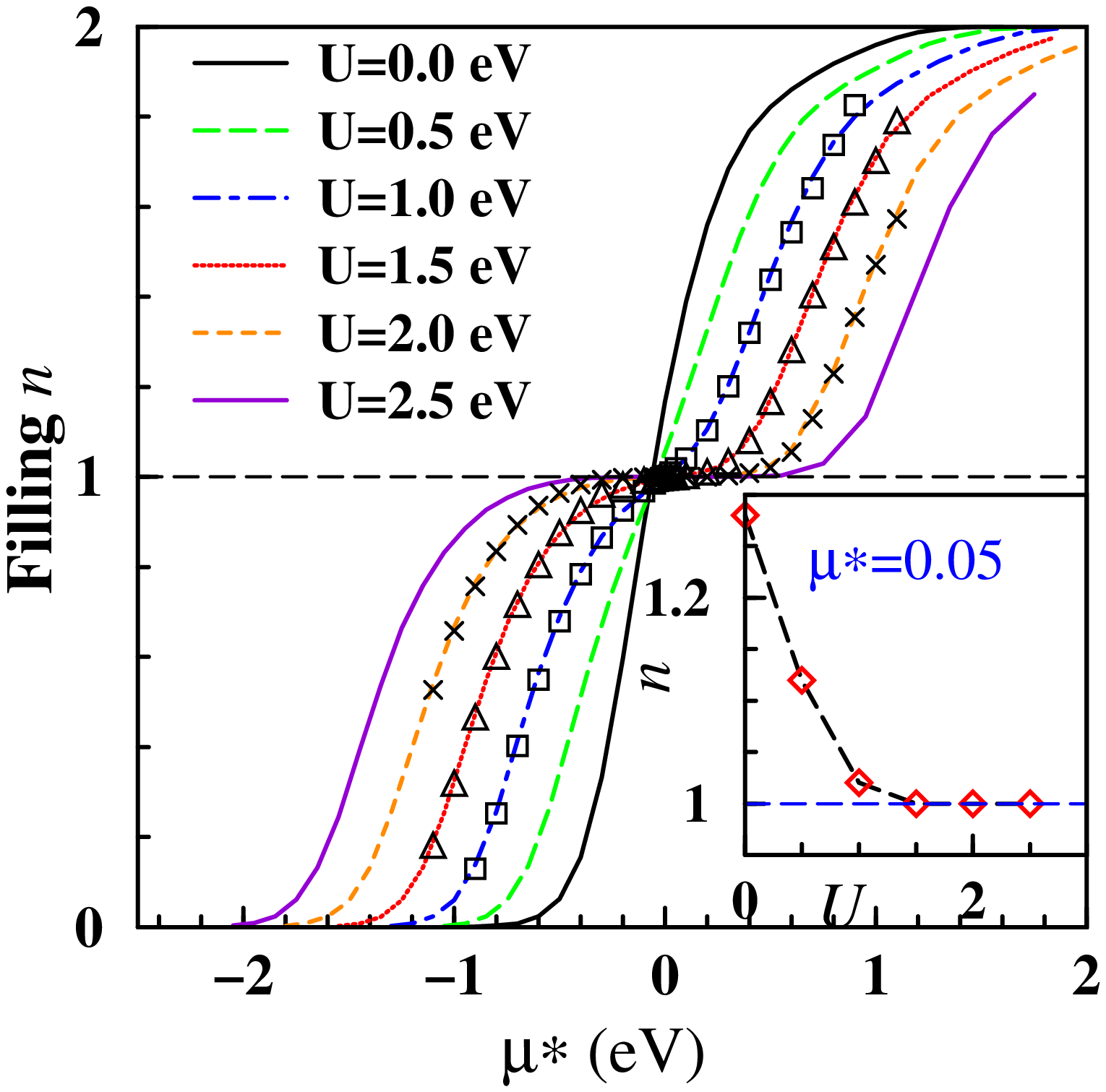}}}
\rotatebox{-90}{\resizebox{6.5cm}{7.5cm}{\includegraphics{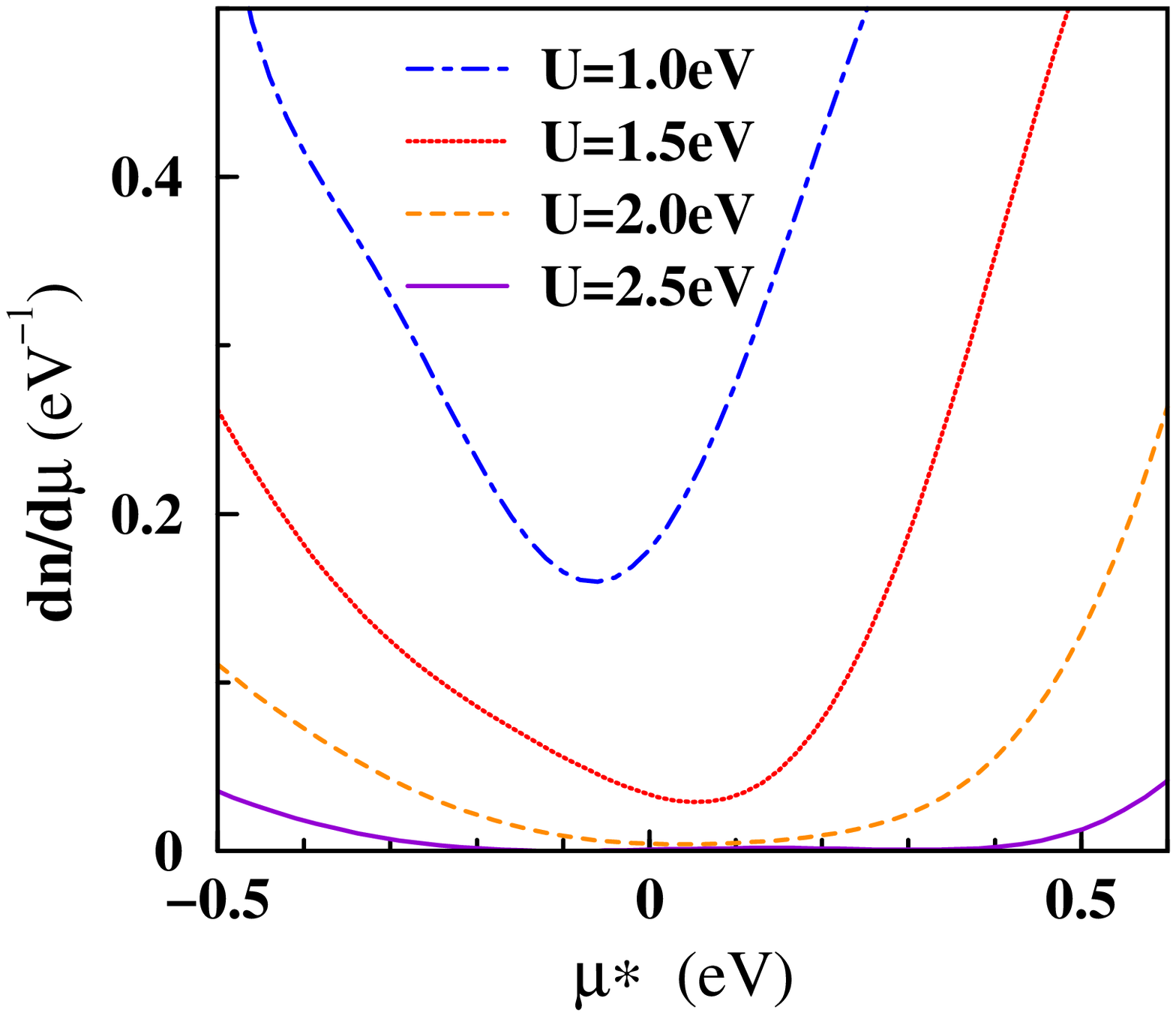}}}
\caption{(Color online) Top: Electron filling $n$ versus
 effective chemical potential $\mu^* = \mu - U/2$, when changing the on-site
 Coulomb repulsion $U$.
The lines and
symbols are the results of DMFT and DQMC calculations respectively.  The
excellent agreement is partly a consequence of the magnetic frustration of
the triangular lattice which diminishes the importance of including
intersite correlations (see text).
 Even for $U$=1.0 eV, the slope changes
 significantly near half-filling, which is indicated
 by the dashed horizontal line.
 Above $U$=1.5 eV, there is a plateau at half-filling.
Inset: $n$ vs. $U$ plot at $\mu^*$=0.05, which displays clearly 
 a metal-to-insulator transition completing ($n\rightarrow$ 1) near $U$=1.5 eV.
 Bottom: Compressibility $\kappa \equiv dn/d\mu$ versus $\mu^*$
 near the critical $U$ regime, showing more clearly where it approaches
and becomes vanishingly small.
}
\label{mun}
\end{figure}

\begin{figure}[tbp]
\rotatebox{-0}{\resizebox{7.5cm}{8.5cm}{\includegraphics{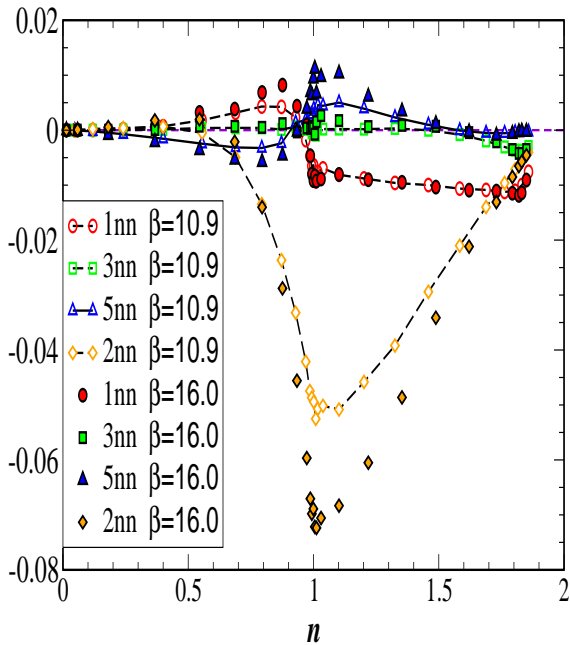}}}
\caption{(Color online)
Determinant QMC result for the spin-spin correlation function
 for temperatures of 1100 K ($\beta$=10.9) and 750 K ($\beta$=16), versus
 versus band filling $n$. $U$=1, and energies are in eV. The 5th neighbor (5nn) is
the third neighbor along a straight line from the reference site.
}
\label{SpinCorr}
\end{figure}

We first address the metal-insulator transition (MIT) at half-filling.
When a gap exists, the occupied electron density remains unchanged 
for values of the chemical potential $\mu$ within the gap. 
This behavior can be observed as a plateau in electron filling $n(\mu)$ plots 
as $U$ is increased, and is displayed in the top panel of Fig. \ref{mun}.
Since there is a Hartree-Fock-like shift of the entire spectrum
with $U$ for the form of Hamiltonian we use, the chemical 
potential is shifted simply by $U/2$ 
and denoted by $\mu^*$.
Already at $U$=1 eV, an inflection is clear near half-filling.
This inflection point becomes a plateau of $\sim$0.1 eV width 
at $U$=1.5 eV, indicating proximity to the MIT.

The critical coupling strength $U_c$ for the MIT can be determined  more 
precisely using two different methods.
First, choosing $\mu^*$=0.05 eV (chosen somewhat above $\mu^*=0$ for
clarity), the filling approaches unity
as $U \rightarrow$1.5 eV, as shown in inset of Fig. \ref{mun} (top panel). 
Second, and most clearly, vanishing compressibility $\kappa=dn/d\mu$ signifies 
the onset of the insulating phase.
The bottom panel of Fig. \ref{mun} shows $\kappa$ is reduced by an 
order of magnitude for $U$=1.5 eV, but only truly vanishes for 
$U$ close to 2 eV.  It must be recalled however that the calculations
are done at T=0.1 eV, and temperature broadening is $\sim\pi$T in addition
to other temperature effects, so
requiring $\kappa$ to vanish overestimates $U_c$.  
Considering the $n(\mu^*)$ plot in the inset of Fig. \ref{mun}, we
identify $U_c\approx 1.5$ eV as the (low temperature) critical strength. 

Figure \ref{SpinCorr} shows the intersite spin correlations
$ <S_i^+S_j^->$, obtained with
DQMC on a 9$\times$9 lattice.  Two temperatures
are shown, $T=1100$K and $T=750$ K.  The dominant tendency
is toward antiferromagnetic orientation between 2nd neighbor spins
connected by $t_2$, which is certainly reasonable given that
it is the largest hopping, and hence has the largest
value of $J=4t^2/U$.  This short-range order increases strongly as $T$
is lowered, and the asymmetry around half-filling becomes sharp and more
pronounced.  Near neighbor spins, connected by $t_1$,
have a correlation which is almost an order of magnitude
smaller, and its increase with lowering temperature is less pronounced.  
This small near neighbor correlation is strongly dependent on band filling,
being ferromagnetic for $n<1$ and
antiferromagnetic for $n>1$.  Interestingly, while
the 3rd neighbor spins connected by $t_3$ are essentially uncorrelated,
spins which are separated linearly by three lattice constants (5th neighbors),
for which there is no direct hopping,
exhibit significant antiferromagnetic correlations for
$n<1$ and ferromagnetic correlations for $n>1$, {\it i.e. opposite}
in sign to the near neighbor correlation.
Overall, the local spin order is markedly different on the
two sides of half-filling, $n=1$, in contrast to the
square lattice (and bipartite lattices generally)
where the order is symmetric about half-filling.

Coupling to second neighbors alone (as occurs here, crudely speaking)
separates the triangular lattice into three independent triangular
sublattices, each with lattice constant $\sqrt{3}a$.  On each sublattice
there are ``near neighbor'' antiferromagnetic correlations, which
are frustrated on the triangular lattice. 
Near neighbor hopping couples a given
site to three sites of each of the other sublattices, and antiferromagnetic
1nn coupling tends to add to the frustration.  Third neighbor hopping
also connects to three sites of each of the other two sublattices, although
the sites are far enough apart that any additional frustration is probably
not an issue.
This picture leads perhaps to even stronger frustration than on the simple
triangular lattice, and rationalizes
why the results for $n(\mu)$ and the compressibility (and perhaps 
other properties) are so similar in DQMC and in DMFT (which neglects
intersite correlations).

\subsection{U-dependent Spectral Density at Half-filling}
\begin{figure}[tbp]
\rotatebox{-90}{\resizebox{6.5cm}{7.5cm}{\includegraphics{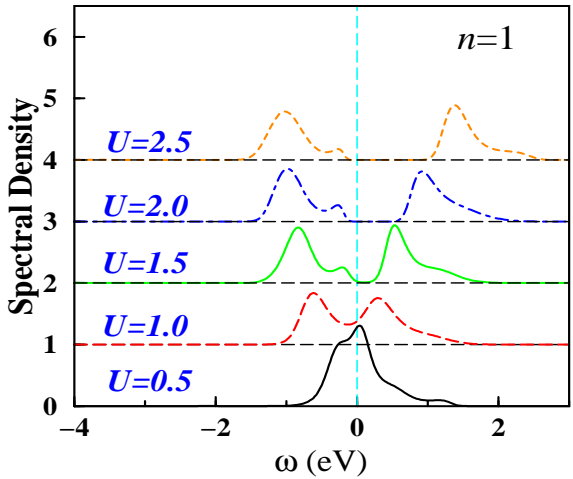}}}
\caption{(Color online) Effect of the on-site Coulomb repulsion $U$
 (in units of eV) on spectral density, which is obtained from MaxEnt,
 for $n$=1.
 At $U$=1.5 eV, a gap opens clearly.
}
\label{sp1}
\end{figure}

At half-filling, the MIT can also be observed 
through the evolution of $A(\omega)$, shown in Fig. \ref{sp1}.
For $U$=1 eV, the spectral density has clearly begun 
splitting into two parts, the lower (LHB) and upper Hubbard bands (UHB).
At $U$=1.5 eV, a gap has opened, consistent with the identification
above of $U_c\approx$ 1.5 eV.  The gap increases monotonically with $U$
for stronger coupling, as expected.  No peak is seen at the chemical
potential as the gap opens because of the relatively high temperature
used here (roughly three times higher than used by Aryanpour and
collaborators\cite{karan} for the case of near neighbor hopping only).

This spectral density shows two interesting features.
First, this gap opening occurs simultaneously with appearance of
a low binding energy peak just below $\omega=0$.
Second, recall that within LDA ($U$=0) half of the weight 
lies in the lower 13\% of the total bandwidth (see Fig. \ref{band}), 
but that at half-filling equal weight goes to the LHB and UHB. 
The band asymmetry fades fairly rapidly as $U$ increases, with
a remnant peak on the low-binding-energy side of the LHB and
the high-binding-energy tail on the UHB  
reflecting the structure in the underlying $U$=0 density of states
(again, see Fig. \ref{band}).

In single band near-neighbor models with a single hopping $t$,
the correlation strength is gauged in terms of the ratio of $U/W$
with an unambiguous bandwidth $W$.
However, in the triangular lattice, the tight-binding DOS has a long
high energy tail and steep lower edge, as shown in Fig. \ref{band}.
This extremely asymmetric shape suggests that $W$ alone does not give
a meaningful measure of the DOS.
We define an effective band width $\widetilde{W}$ in terms of
the second moment of the DOS.
From $\widetilde{W} \approx$0.60 eV we obtain a critical Coulomb
interaction strength ratio $U_c/\widetilde{W}\approx$2.5
for the metal-to-insulator transition.

Using DMFT(QMC), Aryanpour {\it et al.} investigated the half-filled triangular
lattice\cite{karan} with only nearest neighbor hopping ($t=-1$ in
arbitrary units)
for which $\widetilde{W} \approx$4.9 (the full bandwidth is $W=9|t|=9$).
They obtained $U_c$=12$\pm$0.5 eV at a temperature of 400 K, or 
$U_c/\widetilde{W} \approx$ 2.5--2.6,
the same as for Li$_x$NbO$_2$ ($x$=0) above (carried out at 1160 K).
Thus in this case the more complex dispersion and altered DOS shape has
negligible effect on the critical interaction strength ratio.

\subsection{Local Susceptibility at Half-filling}
\begin{figure}[tbp]
\rotatebox{-90}{\resizebox{6.5cm}{7.5cm}{\includegraphics{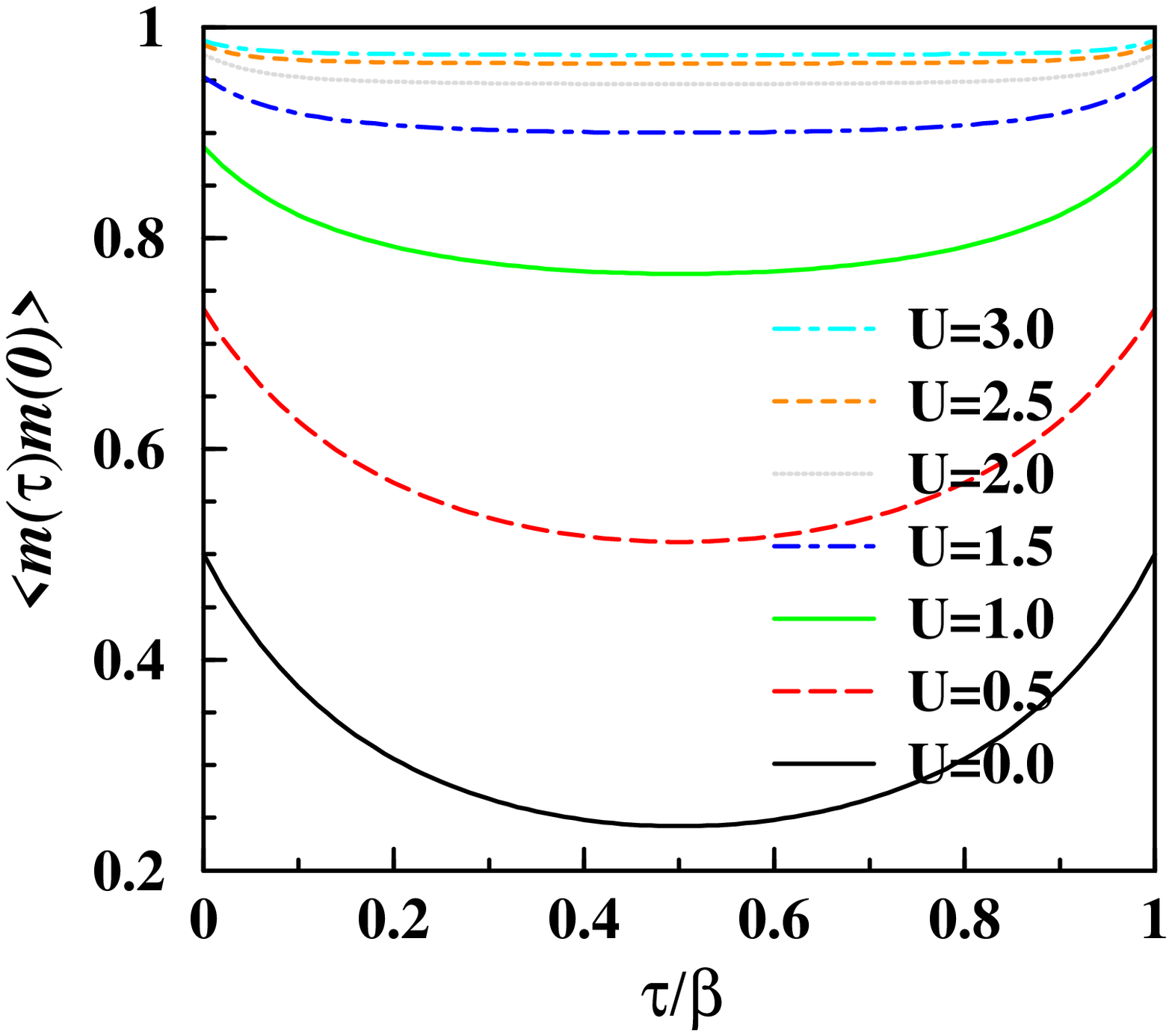}}}
\rotatebox{-90}{\resizebox{6.5cm}{7.5cm}{\includegraphics{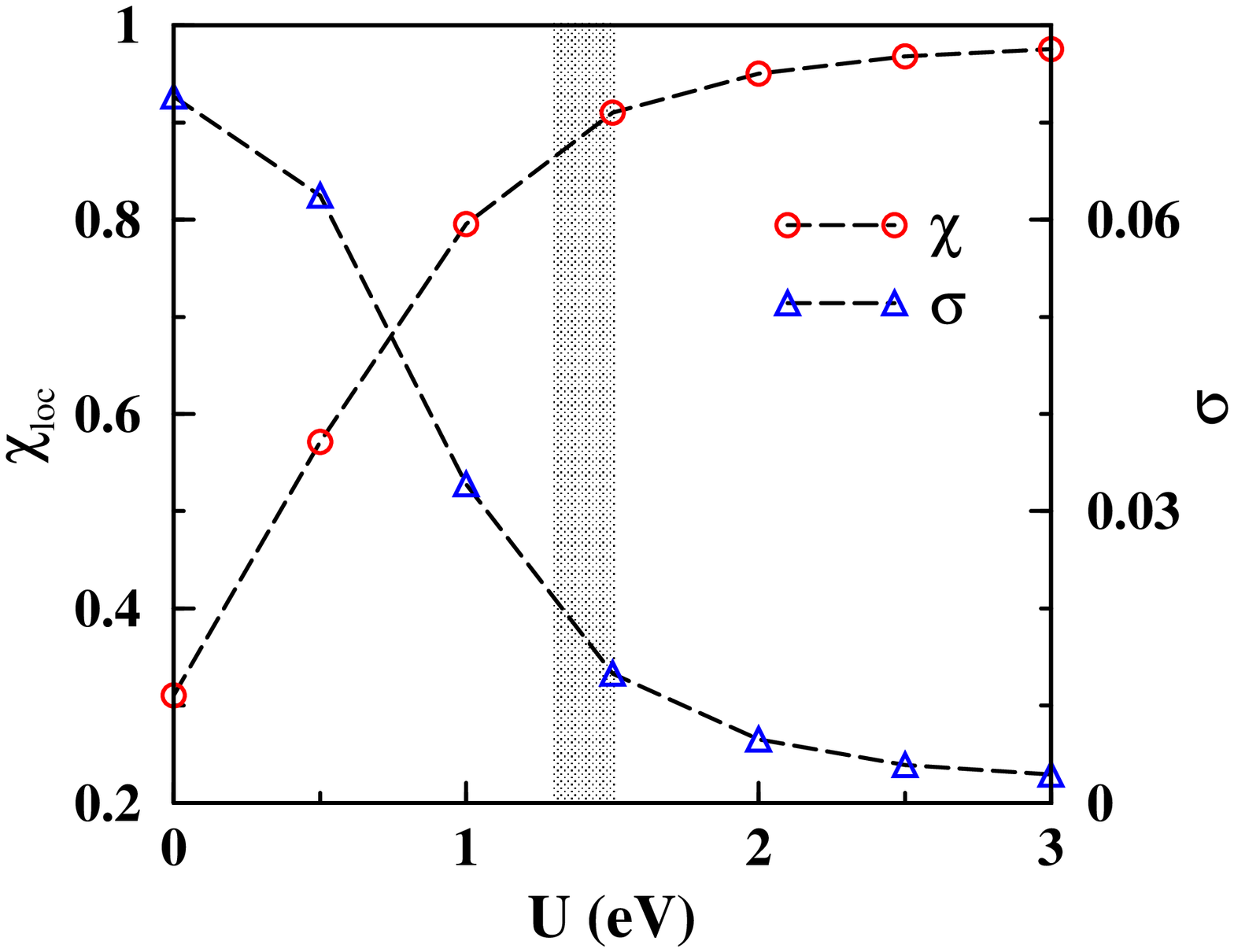}}}
\caption{(Color online) 
 Top: Change in the correlation function $\langle m(\tau)m(0) \rangle$
 of local moment $m$ as $U$ increases.
 Bottom: $U$-dependent local susceptibility $\chi_{loc}$ (left)
 and standard deviation $\sigma$ (right) over (0,$\beta$)
 for the correlation function.
 The $\sigma$ plot shows that variance of the correlation 
 function is strongly reduced by the transition to the insulating phase.
  }
\label{nm}
\end{figure}

At half-filling, the local moment 
$m_{rms} = \sqrt{\langle m_z^2 \rangle} \equiv  
\sqrt{\langle (n_\uparrow -n_\downarrow)^2 \rangle}$
increases from $\frac{1}{\sqrt{2}}$ at $U$=0 to unity
for $U\rightarrow \infty$. For the half-filled system $n$=1,
$m_{rms} = \sqrt{1-2d}$ where $d$ is the fraction
of doubly-occupied sites.
The imaginary time correlation function of the local moment
$\chi(\tau) = \langle m(\tau)m(0) \rangle$ contains additional information
about how the moment decorrelates in imaginary time from its value
$\chi(\tau=0) =\langle m_z^2 \rangle$.
As shown in Fig. \ref{nm}, as $\tau$ increases in the range
[0,$\beta$/2], $\chi(\tau)$
decreases strongly when $U$ is small,
whereas for $U > U_c$, $\chi(\tau)$ becomes nearly flat.

The bottom panel of Fig. \ref{nm} shows the local susceptibility
$\chi_{loc}\equiv (1/\beta) \int_{0}^{\beta} \chi(\tau) d\tau$
versus $U$.  When $U$ surpasses $U_c$, the local susceptibility 
rapidly saturates.
The variance $\sigma$ of $\chi(\tau)$ from its average value ($\chi_{loc}$)
is also displayed in Fig. \ref{nm}. $\sigma(U)$ drops by roughly
a factor of 40 between $U$=0 to $U$=3 eV, having its maximum (negative)
slope around $U$=1.

\section{Nonstoichiometric Phase}
In this section, we will address the hole-doped phases, 
specifically two fillings  $n$=4/3
(Li$_\frac{1}{3}$NbO$_2$, representing the regime not yet synthesized)
and $n$=5/3 (Li$_\frac{2}{3}$NbO$_2$, representing the reported
superconducting regime) through spectral density studies.
Comparison of these results with experimental data will enable
the identification of the strength of correlation effects.

\subsection{Correlation Effects on Spectral Density}
\begin{figure}[tbp]
\rotatebox{-90}{\resizebox{6.5cm}{7.5cm}{\includegraphics{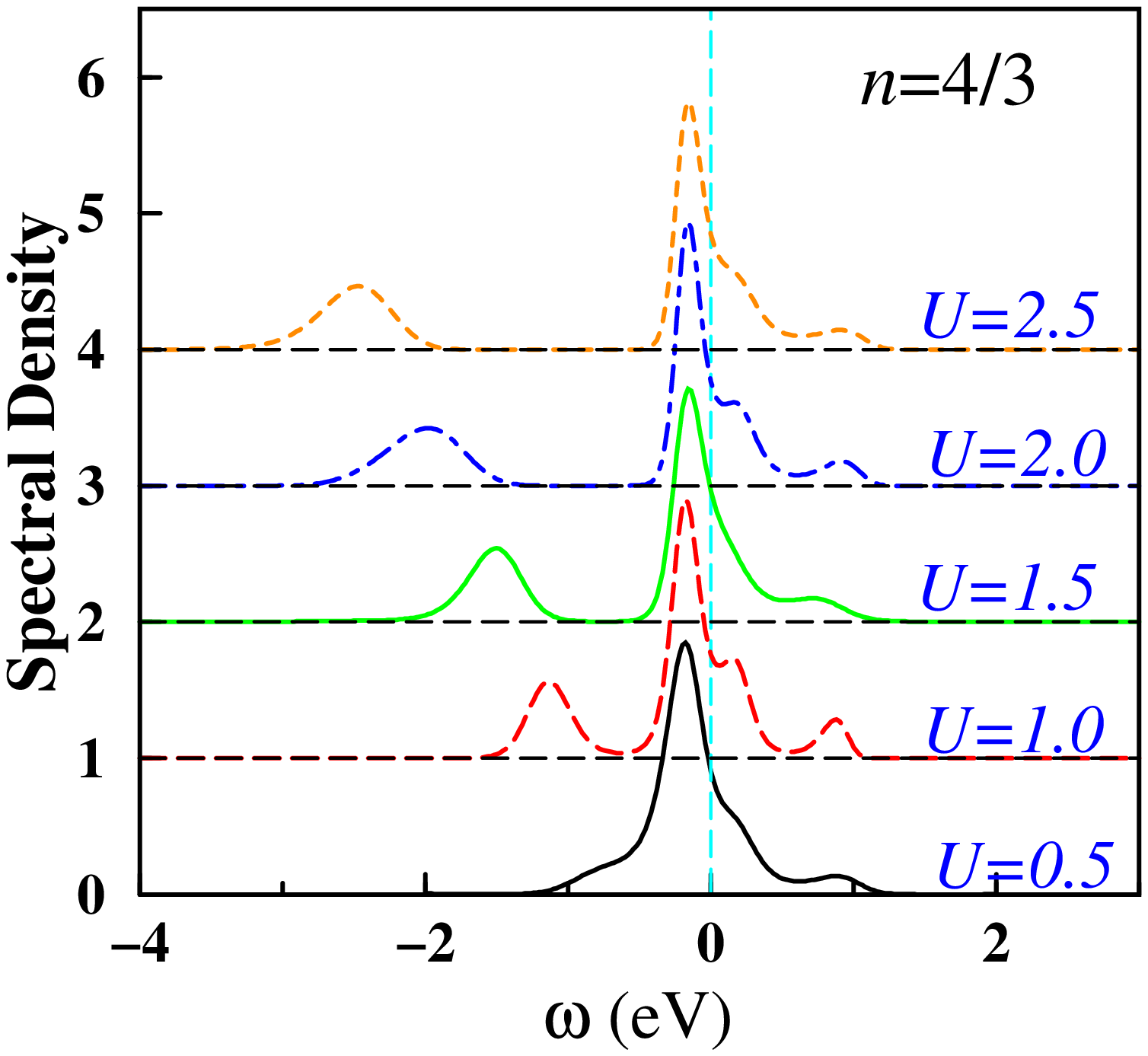}}}
\rotatebox{-90}{\resizebox{6.5cm}{7.5cm}{\includegraphics{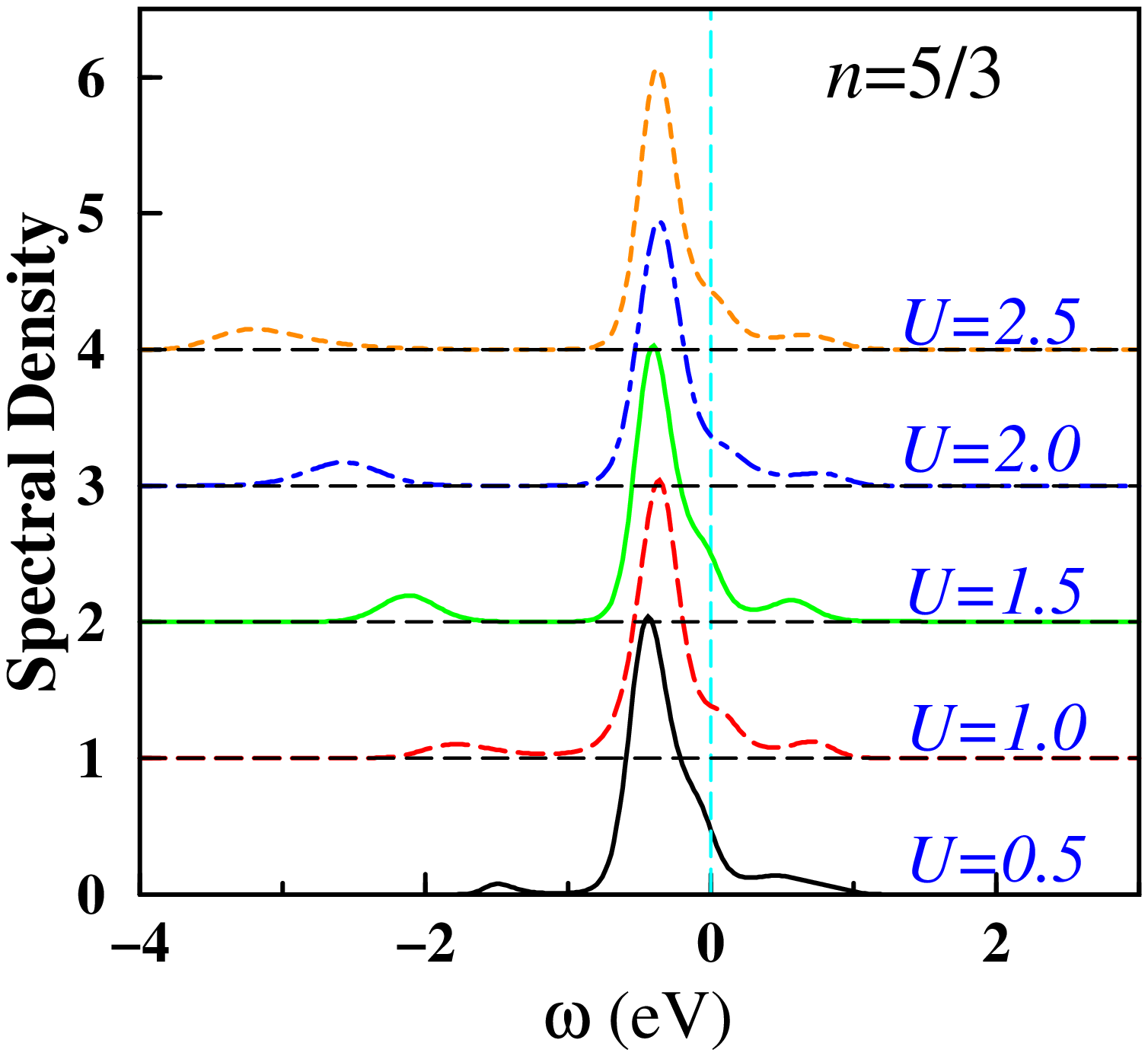}}}
\caption{(Color online) Effect of the on-site Coulomb repulsion $U$
 (in units of eV) on spectral densities, which are obtained from MaxEnt,
 for $n$=4/3 (top) and 5/3 (bottom).
 As expected from the bandwidth,
 correlated behavior already appears at $U$=0.5 eV.
 (For details, see text.)
 Note the spectral densities are normalized, i.e., $\int 
A(\omega)d\omega =1$.
}
\label{sp45}
\end{figure}

Figure \ref{sp45} shows $A(\omega)$ for $n$=4/3 and 5/3.
At $n$=4/3, even for $U$=1.0 eV, a robust lower Hubbard band (LHB) 
is established. 
By $U$=1.5 eV=$U_c$ the LHB completely separates from 
the more intense peak 
at $-0.15$ eV.
The LHB  is clearly identifiable and contains about 0.3 electrons per spin, 
which is close to the number of holes per spin $n_h$=1/3.
The structure around the chemical potential and around 0.5 eV
is persistent as $U$ increases, but is not monotonic and may
not be above the resolution of the MaxEnt technique that is used
to obtain these spectral densities.  Certainly there is no
obvious upper Hubbard band that shifts monotonically with $U$. 

At $n$=5/3, the LHB appears for $U$ 
as small as 0.5 eV.  Its weight is near 0.1 per spin independent of $U$,
and shifts downward almost linearly with $U$; the hole-doping level 
is $n_h$=0.17/spin.  For the main spectral weight peak around -0.5
eV and in the electron-addition spectrum, there is little change
as $U$ increases.
The LHB can be understood simply.  In a snapshot of a doped system
one would see doubly occupied sites and singly occupied sites (for
substantial $U$).  The process corresponding to the LHB is removing
an electron from the singly occupied sites, so the fraction of such
sites is equal to the number of holes, and thus similar to the
weight of the LHB.  This picture seems to be more precise for the
higher hole concentration ($n$=4/3).

Cherkashenko {\it et al.} reported\cite{cherk} the x-ray emission
spectrum for Nb $4d\rightarrow 2p$ transitions for Li$_x$NbO$_2$,
$x$=0.97 and 0.71, {\it i.e.} for the near band insulator and
superconducting compositions respectively.  With their low
resolution, all that could be identified was a decrease in the
occupied Nb $4d$ intensity and some narrowing,
without measurable shift in the peak, as
the Li concentration was reduced.   This behavior is roughly consistent
with a rigid band picture, if the narrowing can be ascribed to the
Fermi level moving from the top of the band (insulating phase) down
into the band (metallic phase).   However, the edge position did not change
as it would in a rigid band picture,
so the interpretation is not certain. 
Higher resolution
x-ray measurements, compared with our predictions, will be very useful
in identifying correlation effects in this system.  

\subsection{Effect of Nb--O Bond Stretching Mode}
\begin{figure}[tbp]
\rotatebox{-90}{\resizebox{6.5cm}{7.5cm}{\includegraphics{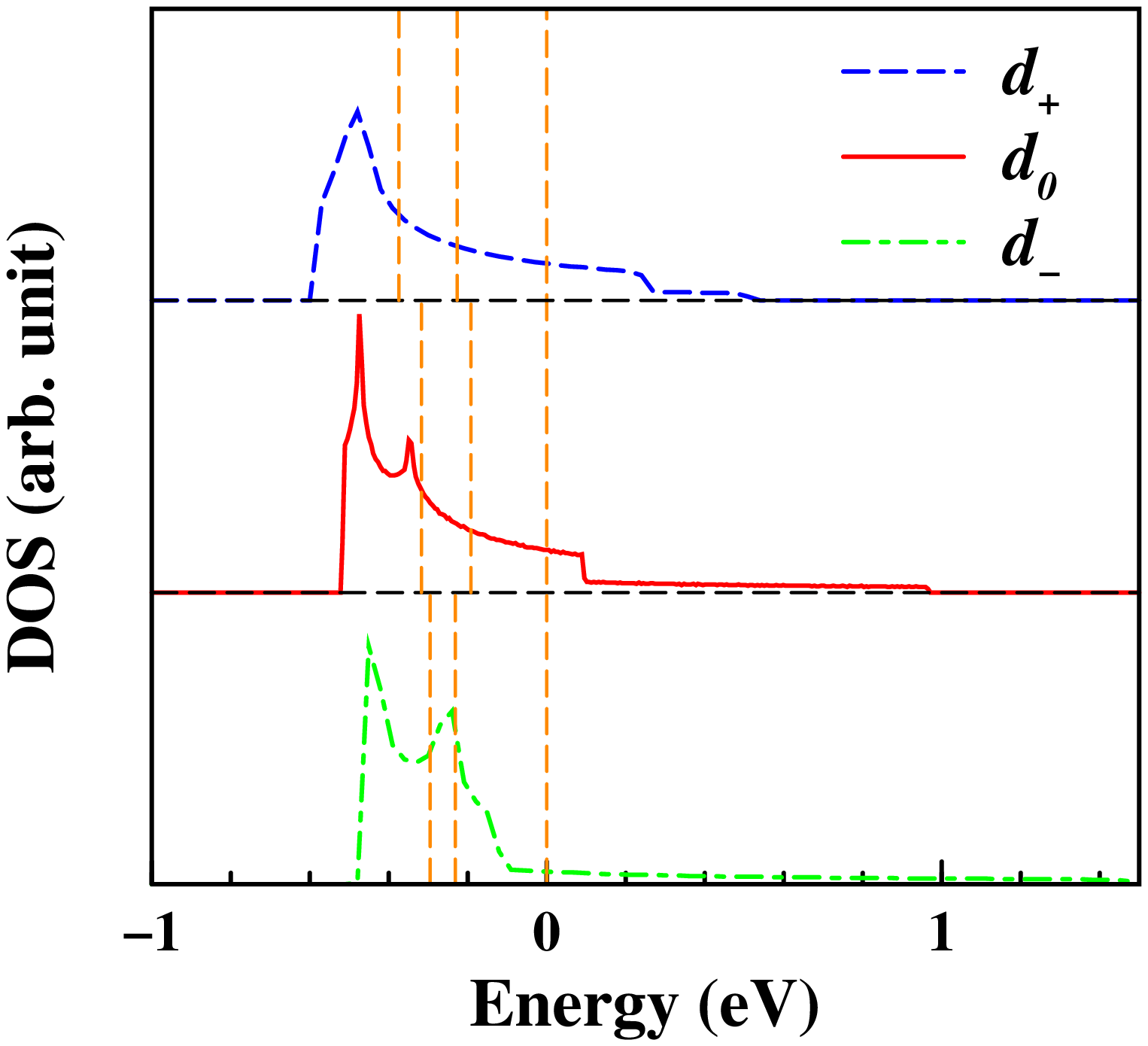}}}
\rotatebox{-90}{\resizebox{6.5cm}{7.5cm}{\includegraphics{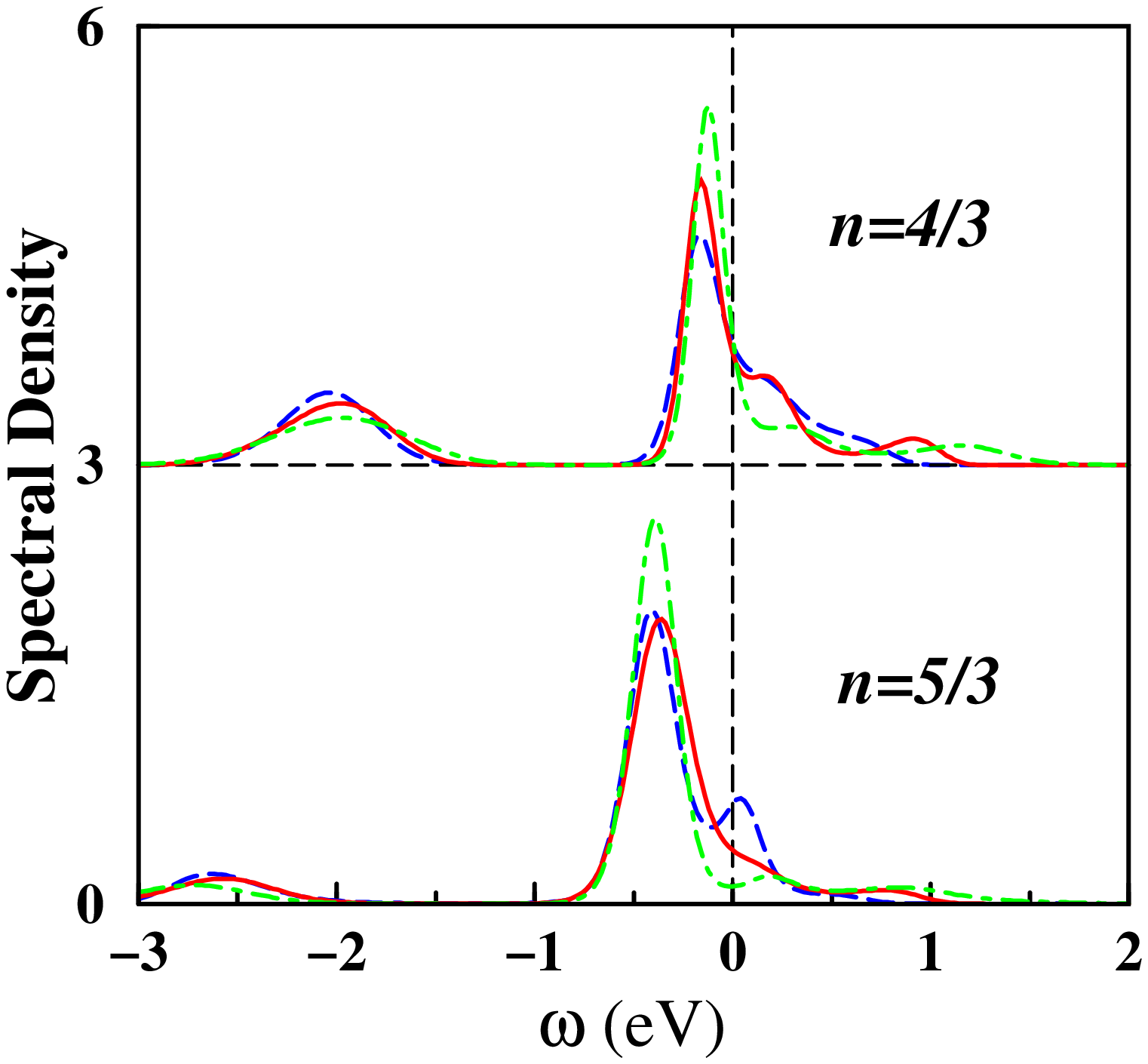}}}
\caption{(Color online) Effect of O displacement on spectral densities.
 O is displaced by $\pm$0.10~\AA~ in the $\langle 001 \rangle$ direction.
 $d_0$ is the optimized position.
 $d_-$ and $d_+$ denote contraction and elongation of the Nb--O bond length,
 respectively.
 Top: Tight-binding DOS (i.e., $U$=0) for the O displacement.
 The densities of states are aligned with respect to $E_F$ for $n$=5/3,
 set to zero. For each case, the dashed vertical lines denote $E_F$
 for $n$=1, 4/3, and 5/3 (left to right).
 Bottom: Change in spectral densities, which shows effects of 
 the phonon mode, at $U$=2 eV.
 The line designations are the same as in the top panel.
}
\label{usp}
\end{figure}

\begin{table}[bt]
\caption{Change in direct width $W$, effective width $\widetilde{W}$
 (see text for the definition),
 and critical $U_c$ for MIT (in units of eV) when O is displaced.  See
 the caption of Fig. \ref{usp} for the definition of the displacements
 $d_+, d_-$.
 The centroid of the band $E_0$ (in meV) measured from the value of $E_F$
 for $n$=4/3.
 There is a consistent critical strength $U_c/\widetilde{W} \approx$2.5.
 Each tight binding parameters (in meV) used here are from Ref. \cite{erik1}.
 The compared DOS are shown in the top panel of Fig. \ref{usp}.
}
\begin{center}
\begin{tabular}{cccccccc}\hline\hline
     &\multicolumn{3}{c}{Tight binding}& & &  \\\cline{2-4}
~~~  &~~ $t_1$ ~~&~~ $t_2$ ~~&~~ $t_3$ ~~&~~$W$ ~~&~~ $\widetilde{W}$ ~~
&~~ $E_0$ ~~&~~ $U_c$ ~~\\\hline
$d_+$ & 19  & 102  & 10 & 1.1 & 0.51 & -48 & 1.2\\
$d_0$ & 64  & 100  & 33 & 1.5 & 0.60 & -18 & 1.5\\
$d_-$ & 122 & 94   & 56 & 2.0 & 0.80 & -77 & 1.7\\
\hline\hline
\end{tabular}
\end{center}
\label{table1}
\end{table}

The calculations of Ylvisaker and Pickett in the band insulating
phase led to a rough scale for the deformation potential ${\cal D}=dW/dz
\sim 2.5$ eV/\AA~(half the bandwidth change per unit displacement),
which is a substantial value that suggests the importance of O displacement 
$\Delta z$ for electron-phonon
coupling, {\it i.e.} transport and superconductivity.\cite{erik1}
To investigate effects of the O $A_g$ phonon mode on $A(\omega)$
we have used O ion displacements
$\Delta z = \pm$0.10~\AA~ along the $c$ direction.
This displacement is representative of the maximum amplitude of the $A_g$ mode,
and it happens to be similar to the variation of the S position in
NbS$_2$ reported 
between $x$=0 and $x$=0.67 in isovalent, isostructural, and 
superconducting Li$_x$NbS$_2$.\cite{salyer,jellinek}
This magnitude of displacement leads to significant changes in 
hopping integrals,
as reproduced in Table \ref{table1}, resulting in considerably 
different DOS shape for each O position.

The DOS (obtained from the tight-binding fit, but equivalent to the
first principles result) displayed in the top panel of Fig. \ref{usp}
shows that the most significant change is in bandwidth, 
about $\pm$0.5 eV for $\mp$0.10~\AA~ change in O position giving the
mean deformation potential mentioned above.
The variation of the DOS is not well characterized simply by W however, because
although W increases, the portion of the DOS containing $\sim60-70$\% of the
states {\it narrows} by roughly a factor of two.
As a result, the position of $E_F$ and DOS at $E_F$ $N(E_F)$ change 
significantly for all fillings.  This effect is especially operative at $n=5/3$ 
filling: the shifting of E$_F$ across the van Hove singularity reflects the
fact that there is a transfer of charge (holes, in this case) between the low
mass hole pocket at $\Gamma$ and the higher mass hole pocket at the two 
$K$ points.  Since the
bandwidth $W$ seems to change more than other identifiable energy structures, 
the scale ${\cal D}$ = 2.5 
eV/\AA~mentioned above is probably an overestimate of actual deformation 
potentials at the Fermi surface.  The fact that O displacement makes the Fermi
level shift across a van Hove singularity indicates that non-adiabatic 
processes are involved in the neighborhood of $n$=5/3 {\it i.e.} in the
reported superconducting regime.  Similar non-adiabatic effects were 
identified\cite{lilia} 
in MgB$_2$, and have not yet been treated properly.

Considering correlation effects within our DMFT approach, O displacement 
affects $A(\omega)$ most strongly within $\pm0.5$ eV of the
chemical potential, as shown in
the bottom panel of Fig. \ref{usp}. 
The LHB is hardly affected, either in position or in weight.
The spectral densities that are displayed are for $U$=2 eV, 
but for all $U$ studied here the phonon effects are similar.
At $n$=5/3, which is in the range of observed superconductivity,
$A(\omega=0)$ (the interacting analog of $N(E_F)$) fluctuates strongly,
while at $n$=4/3 $A(\omega=0)$ varies little.  The changes occur 
for $\omega$ only up to a few tenths of eV.  The stronger changes for
$n=5/3$ may arise from the already very strong change in the underlying
$N(E_F)$, see the top panel of Fig. \ref{usp}.

\section{Summary}
The single $d_{z^2}$ band in Li$_x$NbO$_2$, isolated 
due to large trigonal crystal
field splitting, requires three neighbor
hopping parameters to reproduce its dispersion, second neighbor hopping
being dominant.  This long range of hybridization reflects the fact that the
corresponding Wannier function is not well localized, having significant
O $2p$ and neighboring Nb $4d$ character in addition to the on-site
$4d_{z^2}$ part. 
The dispersion leads to highly asymmetric DOS which contains half its weight
in the lower 13\% of bandwidth.

Using DMFT(QMC), we have investigated correlation effects on the spectral
densities and have followed the Mott transition using both the compressibility
and the spectral density. At half-filling, a metal-to-insulator transition
occurs at $U_c \approx 1.5$ eV.
Because of long tail and steep edge of DOS in the two-dimensional 
tight-banding model for the triangular lattice, however, we have
introduced an effective bandwidth $\widetilde{W}$ obtained 
from the second moment of the density of states to characterize the band. 
The critical interaction strength ratio
$U_c/\widetilde{W} \sim$ 2.5 for the Mott transition is the same as
was obtained for the triangular lattice nearest neighbor Hubbard model,
indicating that the more complex dispersion does not affect the 
position of the metal-insulator transition in this system.  

Calculation of the spin
correlation functions with determinant QMC verifies that second neighbor
correlations are by far the largest, as expected from the hopping
parameters.  Spin correlations between other neighbors are very minor.
We find however that the behavior of the band filling $n(\mu)$ 
and the compressibility
are virtually identical in DMFT (which neglects intersite spin 
correlations) and DQMC,
justifying {\it a posteriori} our use of DMFT for this triangular lattice
system.  The dominant second neighbor hopping $t_2$ tends to partition
the lattice into three weakly coupled triangular sublattices (lattice
constant $\sqrt{3}$); each of these are frustrated with the antiferromagnetic
coupling, and the small coupling between these sublattices adds to the
frustration.

The spectral density was used to assess the strength and character of
correlation effects in Li$_x$NbO$_2$, both for the static lattice and for
frozen displacements of the O A$_g$ phonon.  These results were presented
for a grid of interaction ($U$) strengths, since the appropriate value for
Li$_x$NbO$_2$ is not yet known.  These results will be useful
in the interpretation of high-resolution photoemission data, which are
needed to further our understanding of this very interesting new triangular
lattice superconductor.   

\section{Acknowledgment}
We acknowledge J. L. Luo for clarifying his specific heat measurements
and K. Aryanpour, S. Chiesa, and E. R. Ylvisaker for useful discussions.
This work was supported by DOE grant No. DE-FG03-01ER45876, by Stewardship
Science Academic Alliance Program (DOE) grant DE-FG01-06NA26204,
and by DOE's Computational Materials Science Network.


\begin{thebibliography}{10}
\bibitem{gese1} M. J. Geselbracht, T. J. Richardson, and A. M. Stacy,
  Nature {\bf 345}, 324 (1990).

\bibitem{bordet} P. Bordet, E. Moshopoulou, S. Liesert,
  and J. J. Capponi, Physica C {\bf 235-240}, 745 (1994).

\bibitem{mosh}  E. G. Moshopoulou, P. Bordet, and J. J. Capponi,
   Phys. Rev. B {\bf 59}, 9590 (1999).

\bibitem{liu} G. T. Liu, J. L. Luo, Z. Li, Y. Q. Guo, N. L. Wang,
  D. Jin, and T. Xiang, Phys. Rev. B {\bf 74}, 012504 (2006).

\bibitem{keller1} D. G. Kellerman, G. P. Shveikin, A. P. Tyutyunnik,
  V. G. Zubkov, V. A. Perelyaev, T. V. D'yachkova, 
  N. I. Kadyrova, A. S. Fedyukov, S. A. Turzhevskii, V. A. Gubanov,
  G. P. Shveikin, A. E. Ka\'rkin, and V. I. Voronin, 
  Superconductivity:Phys. Chem. Tech. {\bf 5}, 2035 (1992).

\bibitem{erik1} E. R. Ylvisaker and W. E. Pickett,
   Phys. Rev. B {\bf 74}, 075104 (2006).

\bibitem{erik2} E. R. Ylvisaker, K.-W. Lee, and W. E. Pickett,
  Physica B {\bf 383}, 63 (2006).

\bibitem{freeman} D. L. Novikov, V. A. Gubanov, V. G. Zubkov,
  and A. J. Freeman, Phys. Rev. B {\bf 49}, 15830 (1994).

\bibitem{keller2} D. G. Kellerman, V. G. Zubkov, A. P. Tyutyunnik,
  V. S. Gorshkov, V. A. Perelyaev, G. P. Shveikin,
  S. A. Turzhevskii, V. A. Gubanov, and A. E. Ka\'rkin,
  Superconductivity:Phys. Chem. Tech. {\bf 5}, 966 (1992).

\bibitem{keller3} A. P. Tyutyunnik, V. G. Zubkov, D. G. Kellerman,
 V. A. Perelyaev, A. E. Ka\'rkin, and G. Svensson,
 Eur. J. Solid State Inorg. Chem. {\bf 33}, 53 (1996).

\bibitem{kumada2} N. Kumada, S. Watauchi, I. Tanaka, and N. Kinomura,
 Mater. Res. Bull. {\bf 35}, 1743 (2000).

\bibitem{tise2} E. Morosan, H. W. Zandbergen, B. S. Dennis,
 J. W. G. Bos, Y. Onose, T. Klimczuk, A. P. Ramirez, N. P. Ong,
 and R. J. Cava,
 Nature Phys. {\bf 2}, 544 (2006).

\bibitem{nacoo} K.-W. Lee, J. Kune\v{s}, and W. E. Pickett,
 Phys. Rev. B {\bf 70}, 045104 (2004); K.-W. Lee, J. Kune\v{s},
 P. Novak, and W. E. Pickett, Phys. Rev. Lett. {\bf 94}, 026403 (2005);
 K.-W. Lee and W. E. Pickett, {\it ibid.} {\bf 96}, 096403 (2006).

\bibitem{nbo2} V. Eyert, Europhys. Lett. {\bf 58}, 851 (2002);
 references therein.

\bibitem{dmft} For review, see A. Georges, G. Kotliar, W. Krauth, and
  M. J. Rozenberg, Rev. Mod. Phys. {\bf 68}, 13 (1996);
  G. Kotliar, S. Y. Savrasov, K. Haule, V. S. Oudovenko,
  P. Parcollect, and C. A. Marianetti, {\it ibid.} {\bf 78}, 865 (2006).

\bibitem{karan} K. Aryanpour, W. E. Pickett, and R. T. Scalettar,
  Phys. Rev. B {\bf 74}, 085117 (2006).

\bibitem{kumada} N. Kumada, S. Muramatu, F. Muto, N. Kinomura,
  S. Kikkawa, and M. Koizumi,
  J. Solid State Chem. {\bf 73}, 33 (1988).

\bibitem{meyer} G. Meyer and R. Hoppe, 
  Angew. Chem., Int. Ed. Engl. {\bf 13} (1974);
  J. Less-Common Met. {\bf 46}, 55 (1976).

\bibitem{gese2}  M. J. Geselbracht, A. M. Stacy, A. R. Garcia, 
    B. G. Silbernagel, and G. H. Kwei, J. Phys. Chem. {\bf 97}, 7102 (1993).

\bibitem{cherk} V. M. Cherkashenko, M. A. Korotin, V. I. Anisimov,
  V. V. Shumilov, V. R. Galakhov, D. G. Kellerman, V. G. Zubkov,
  and E. Z. Kurmaev, Z. Phys. B {\bf 93}, 417 (1994).

\bibitem{fplo} K. Koepernik and H. Eschrig, Phys. Rev. B {\bf 59}, 1743 (1999).

\bibitem{maxent} M. Jarrell and J. E. Gubernatis, 
   Phys. Rep. {\bf 269}, 133 (1996).

\bibitem{bss}
R. Blankenbecler, D. J. Scalapino, and R. L. Sugar
Phys. Rev. D {\bf 24}, 2278 (1981).

\bibitem{salyer} P. A. Salyer, M. G. Barker, A. J. Blake, D. H. Gregory,
 and C. Wilson, Acta Crystallogr. C{\bf 59}, i4 (2003).

\bibitem{jellinek} F. Jellinek, G. Brauer, and H. M\"{u}ller,
 Nature (London) {\bf 185}, 376 (1960).

\bibitem{lilia}L. Boeri,
  G. B. Bachelet, E. Cappelluti, and L. Pietronero,
  Phys. Rev. B {\bf 65}, 214501 (2002).


\end{thebibliography}
\end{document}